\documentclass[prl,aps,letterpaper,floatfix,superscriptaddress,nofootinbib,preprintnumbers,twocolumn,10pt]{revtex4-1}

\usepackage{dcolumn}
\usepackage{bm}
\usepackage[dvips]{graphicx}
\usepackage{epsfig}
\usepackage{slashed}
\usepackage{feynmp}
\usepackage{hhline}
\usepackage{amsmath, amsfonts, amssymb}
\usepackage[normalem]{ulem}
\newcommand{\be}{\begin{equation}}
\newcommand{\ee}{\end{equation}}
\newcommand{\bee}{\begin{equation*}}
\newcommand{\eee}{\end{equation*}}
\newcommand{\bea}{\begin{eqnarray}}
\newcommand{\eea}{\end{eqnarray}}
\newcommand{\bean}{\begin{eqnarray*}}
\newcommand{\eean}{\end{eqnarray*}}

\usepackage{color}

\begin{document}

\title{A Boost to $h \to Z \gamma$: from LHC to Future $e^+ e^-$ Colliders}

\author{Jose Miguel No} 
\affiliation{Department of Physics, King's College London, Strand, WC2R 2LS London, UK}
\affiliation{Department of Physics and Astronomy, University of Sussex, 
Brighton BN1 9QH, UK}

\author{Michael Spannowsky} 
\affiliation{Institute of Particle Physics Phenomenology, Physics Department, Durham University,
Durham DH1 3LE, UK}

\preprint{KCL-PH-TH/2016-71, DCPT/16/222, IPPP/16/111}

\date{\today}

\begin{abstract}
A precise measurement of the Higgs $h \to Z \gamma$ decay is very challenging at the LHC, due to the very low 
branching fraction and the shortage of kinematic handles to suppress the large SM $Z \gamma$ background. 
We show how such a measurement would be significantly improved by considering Higgs production in association with a hard jet. 
We compare the prospective HL-LHC sensitivity in this channel with other Higgs production modes where $h$ is fairly boosted, e.g.~weak boson fusion, and also 
to the potential $h \to Z \gamma$ measurement achievable with a future $e^{+}e^{-}$ circular collider ({\sl fcc-ee}). Finally, 
we discuss new physics implications of a precision measurement of $h \to Z \gamma$.
\end{abstract}

\maketitle


%
%

The $h\to Z\gamma$ decay of the Higgs boson constitutes a challenging collider measurement~\cite{Gainer:2011aa}.
Current LHC constraints from ATLAS and CMS with $7-8$ TeV Run 1 data are very weak~\cite{Chatrchyan:2013vaa,Aad:2014fia}, and
projections for the HL-LHC with an integrated luminosity $\mathcal{L} = 3000\,\,\mathrm{fb}^{-1}$ yield a relative 
uncertainty on the $h \to Z\gamma$ signal strength of at least $\Delta\mu/\mu \sim 0.3$~\cite{ATL-PHYS-PUB-2014-006,ATL-PHYS-PUB-2014-016}.
This is much worse than the projected precision in $h \to W^{+}W^{-},\,ZZ,\,\gamma\gamma$, 
and also significantly less precise than projections for 
the fermionic Higgs decays $h \to b\bar{b},\,\tau^{+}\tau^{-},\,\mu^{+}\mu^{-}$. 

The main reason behind this poor sensitivity is the very low Higgs branching ratio (BR) in the Standard Model (SM)
$\mathrm{BR}(h \to \ell^{+} \ell^{-} \gamma) \simeq 10^{-4}$, together with the 
shortage of kinematic handles in an inclusive search to efficiently suppress the large SM $(Z \to \ell^{+} \ell^{-}) \gamma$ background.
Consequently, attempts to measure $\mathrm{BR}(h \to Z \gamma)$ in inclusive Higgs production (i.e.~without relying on 
the presence of additional jets) suffer from a small signal-to-background
ratio $S/B < 10^{-2}$~\cite{ATL-PHYS-PUB-2014-006} and have to rely on a very large integrated 
luminosity, as foreseen for the HL-LHC, to allow for data-driven background estimates that ameliorate the effect of 
systematic background uncertainties in presence of small $S/B$. 


In this letter we show that it is possible to significantly improve on the precision for this measurement at the 
LHC by considering Higgs production in association with a hard jet. We detail the kinematical advantages this entails in terms 
of signal to background discrimination, and compare the projected sensitivity with the one obtained 
via other Higgs production modes, particularly weak boson fusion (WBF).
We further compare the achievable precision in determining $\mathrm{BR}(h \to Z \gamma)$  
at the HL-LHC with that of a future $e^+e^-$ circular collider (hereinafter \textsl{fcc-ee}), 
which would benefit from a very large amount of integrated luminosity while on the other hand
providing a smaller $e^+e^- \to Z h$ Higgs production cross section.

Finally, we discuss potential implications of a precision $h \to Z \gamma$ measurement for new physics beyond the SM. 
In particular, we show the potential gain of studying the decays $h \to \gamma\gamma$  and $h \to Z \gamma$ in concert to probe 
the quantum numbers of new charged particles that couple to the Higgs boson, in a rather model-independent way.   

\vspace{-3mm}

\subsection{Boosting $h \to Z \gamma$ with Jets at the LHC}
\label{sec:hjet}

\vspace{-3mm}

\begin{center}
\textit{(i) Higgs production in association with a hard jet}
\end{center}

We here focus on the production process $pp \to h\,j$ with the subsequent Higgs decay $h \to (Z \to \ell^+\ell^-) \gamma$, aiming 
to exploit the characteristic decay pattern of a resonance to separate the small signal from the large SM background. 
As the signature consists of a jet, two charged leptons and a photon, the entire final state can be 
reconstructed using objects with small fake rates. Thus, reducible backgrounds are rendered negligible and the only 
relevant SM background process to consider is $Z \gamma$+jets~\cite{Chatrchyan:2013vaa,Aad:2014fia}.

The production of $h$ in association with a boosted jet in the present context provides two major kinematical 
advantages as compared to the inclusive search:~\textit{(i)} While for the signal the invariant mass 
$m_{\ell\ell\gamma}$ distribution peaks around $m_h \simeq 125$ GeV irrespectively of the transverse momentum 
of the extra jet $p_{T}^{j}$, for the background the two variables are correlated and $m_{\ell\ell\gamma}$ 
shifts to larger values as $p_{T}^{j}$ increases, moving away from the signal.~\textit{(ii)}  
A rather soft photon is expected for the SM background as it dominantly comes from initial state radiation. 
This is mildly independent of the value of $p^{j}_T$. In contrast, for the signal the values of 
$p^{j}_{T}$ and $p^{\gamma}_{T}$ are highly correlated, as $\gamma$ inherits part of the large Higgs boost
in events with a hard jet. These two effects, highlighted in Figure~\ref{fig:hboson_ptj}, allow to significantly improve the sensitivity
to $h \to Z \gamma$, as shown below.

We generate the SM signal and background using {\sc MadGraph\_aMC@NLO}~\cite{Alwall:2014hca} including finite top mass effects and 
showering the partonic process with {\sc Pythia} 8~\cite{Sjostrand:2007gs}. Both signal and background are normalized to their
NLO cross section values using a flat $k$-factor, $k = 2.0$ for the signal~\cite{Caola:2016upw} (see also~\cite{deFlorian:2016spz})
and $k = 1.5$ for the background~\cite{Campbell:2012ft}.
For event selection we require two isolated leptons with $p^{\ell}_T > 10$ GeV, $\left| y_{\ell} \right| < 2.5$ and one isolated
photon with $p^{\gamma}_T > 20$ GeV, $\left| y_{\gamma} \right| < 2.5$. Leptons (photons) are considered isolated if the hadronic energy deposit 
within a cone of size $R=0.3$ is smaller than $10\%$ of the $p_T$ of the lepton (photon) candidate. 
Jets are defined using the anti-$kT$ recombination algorithm with $R=0.4$, 
with $p^{j}_{T} \geq$ $50$ GeV and $|y_j| \leq 4.5$.

With two charged isolated leptons and a photon the final state has a high trigger efficiency. 
To reconstruct the Higgs and $Z$ bosons we further require
$
80~\mathrm{GeV} \leq m_{\ell\ell} \leq 100~\mathrm{GeV}
$
and
$
115~\mathrm{GeV} \leq m_{\ell\ell\gamma} \leq 135~\mathrm{GeV}.
$
As the invariant masses in the case of the signal are insensitive to the boost of $h$ and $Z$ bosons, analyzing the effect of an 
increasing jet $p_{T}^{j}$ on the the shape of the $m_{\ell\ell\gamma}$ background distribution will help 
improving the signal-to-background ratio $S/B$.
We perform a varying selection on $p^{j_1}_T$ ($j_1$ being the leading jet in transverse momentum in the event), requiring $p^{j_1}_T > p^{j_{cut}}_T $ with 
three reference $p^{j_{cut}}_T$ values $p^{j_{cut}}_T = 50,\,100$, $180$ GeV.
After selection cuts, these yield rather small event rates, as shown in Table~\ref{Table1}. 
We observe that increasing the value of $p^{j_{cut}}_T$ from $50$ to $180$ GeV improves $S/B$ from 
$0.7\%$ to $2.3\%$, but at the expense of reducing the statistical significance measured in $S/\sqrt{B}$.

\begin{widetext}
  \onecolumngrid
   \begin{figure}[h!]
\centering
\includegraphics[width=0.325\textwidth]{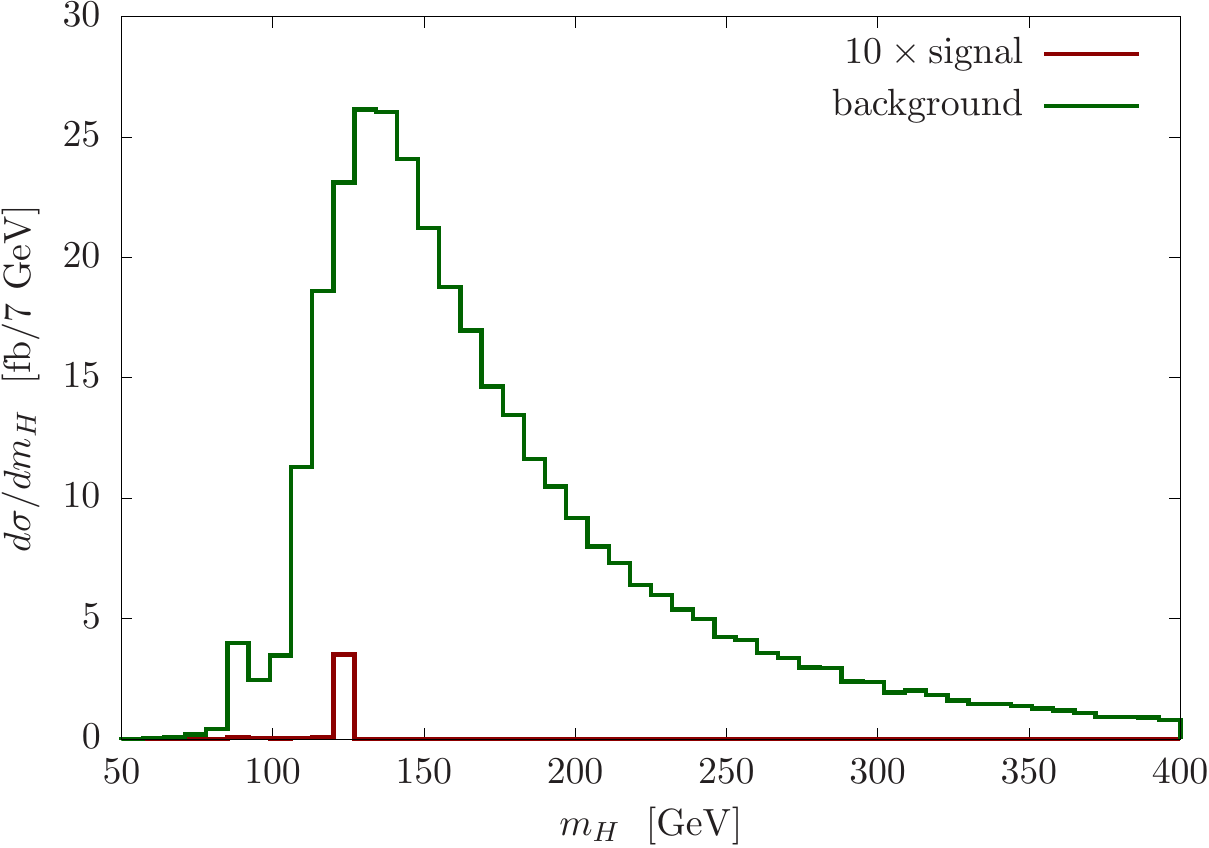}
\includegraphics[width=0.325\textwidth]{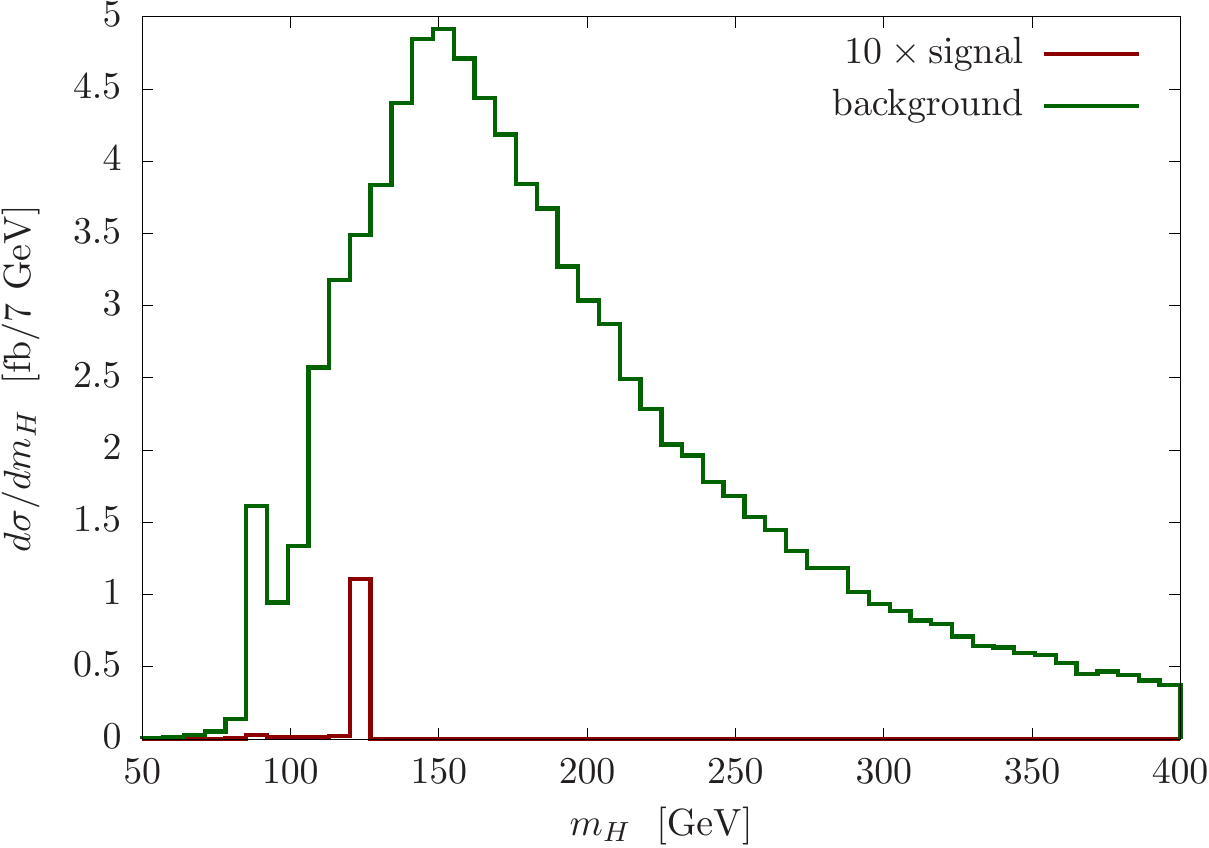} 
\includegraphics[width=0.325\textwidth]{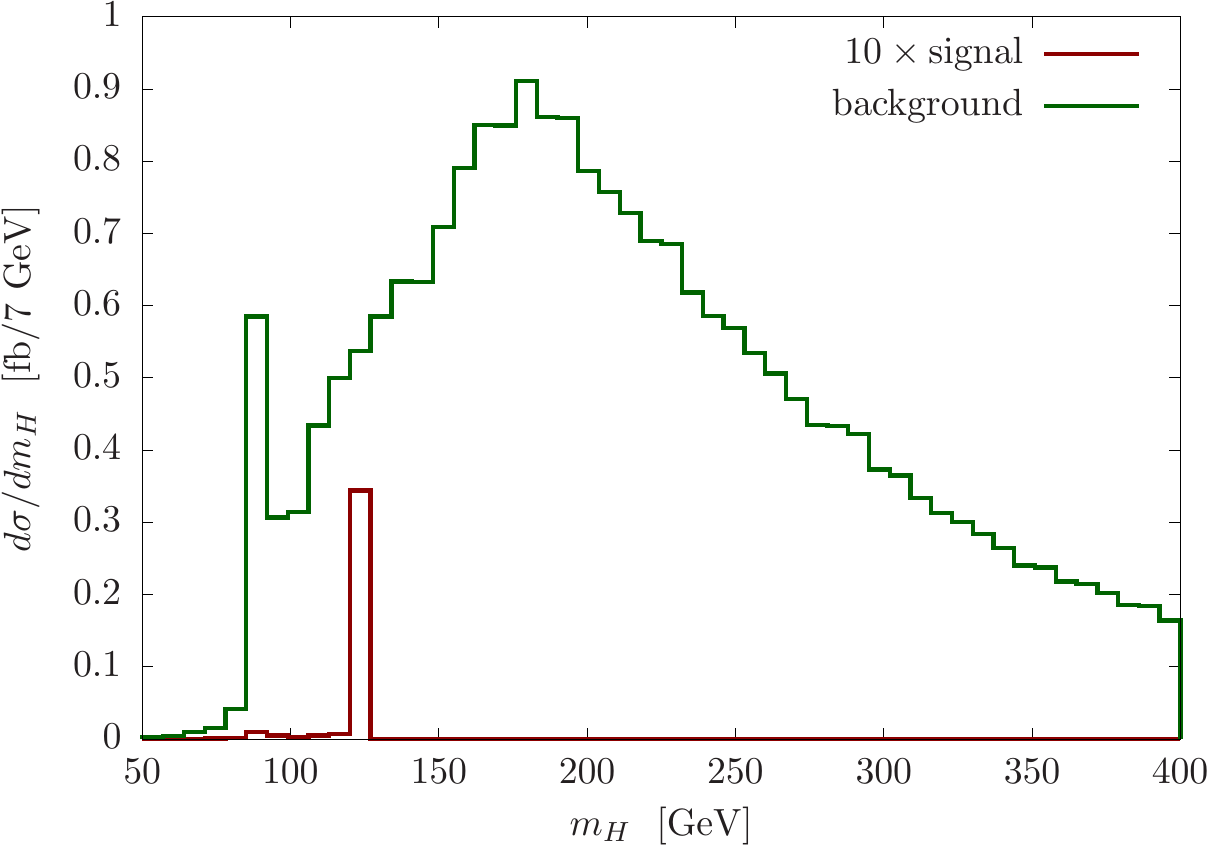} \\
 \includegraphics[width=.325\textwidth]{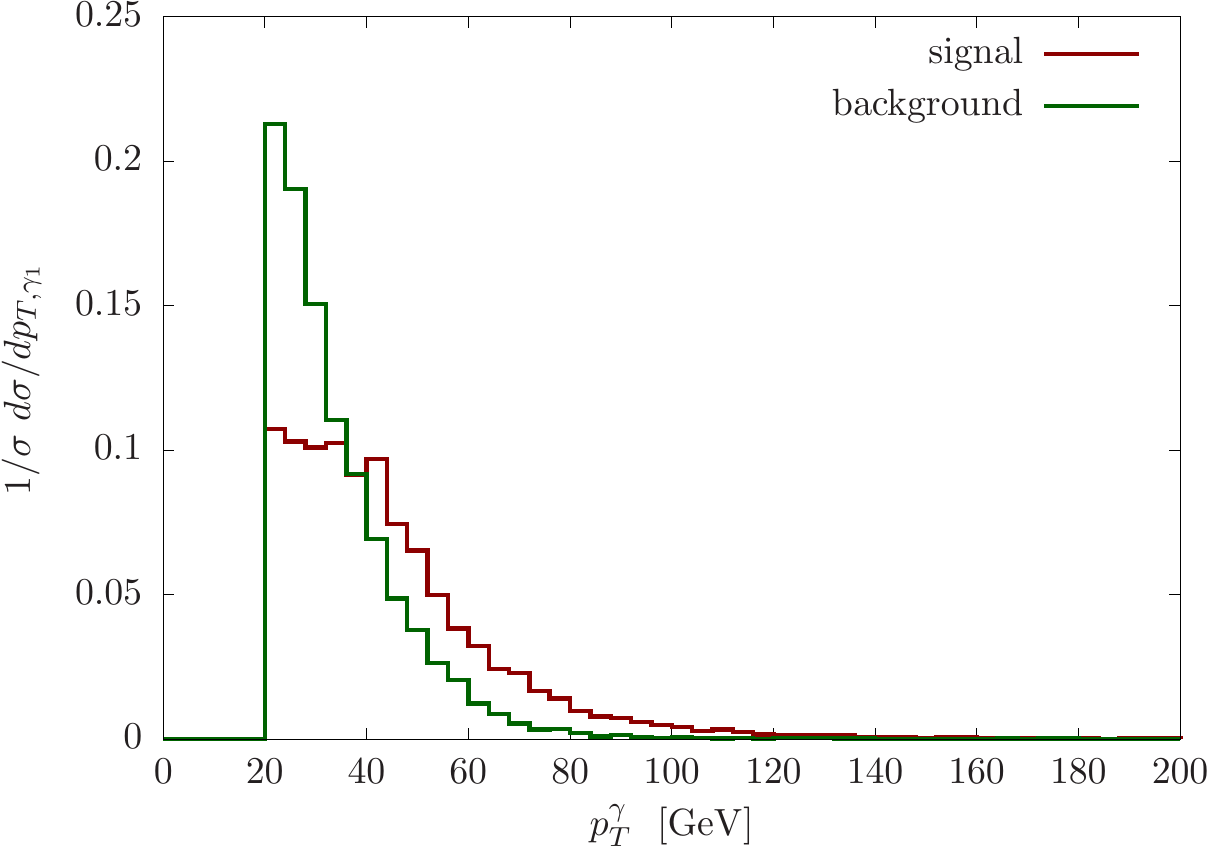}
 \includegraphics[width=.325\textwidth]{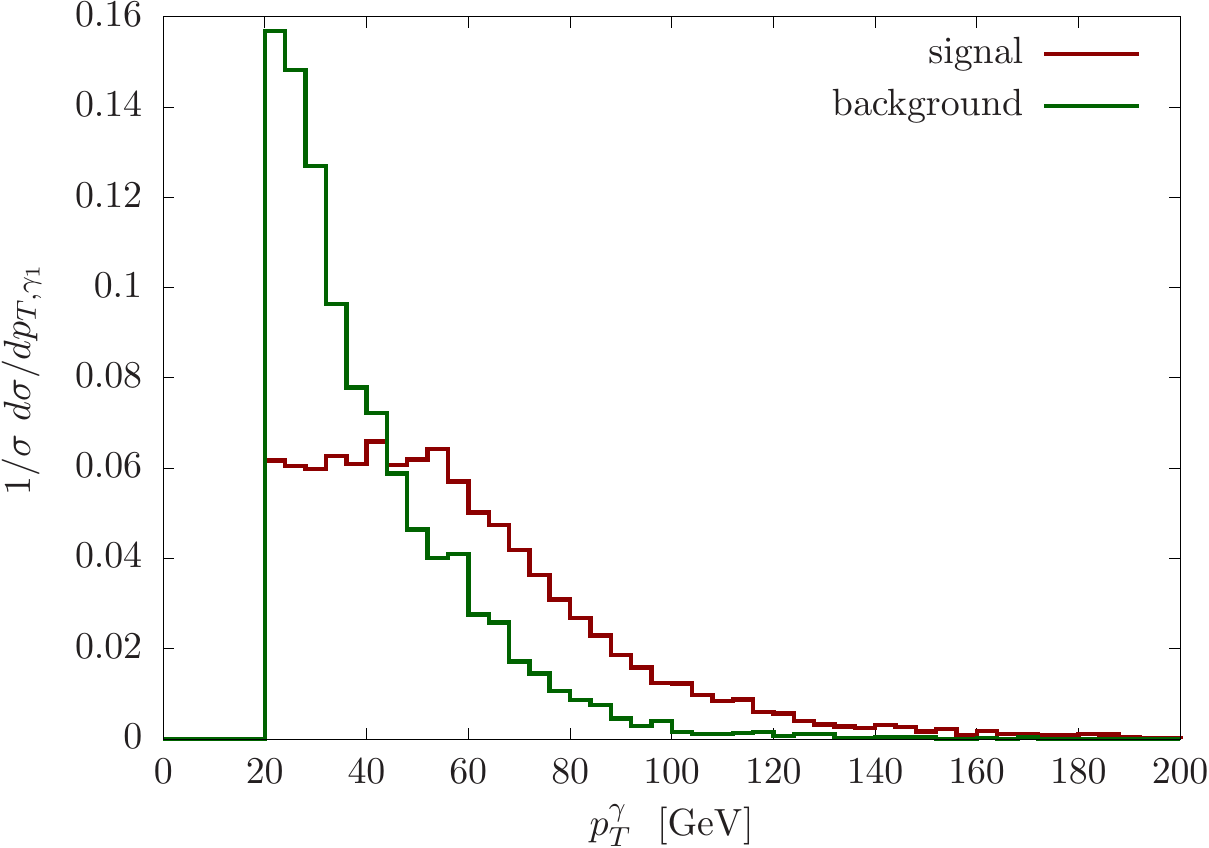}
 \includegraphics[width=.325\textwidth]{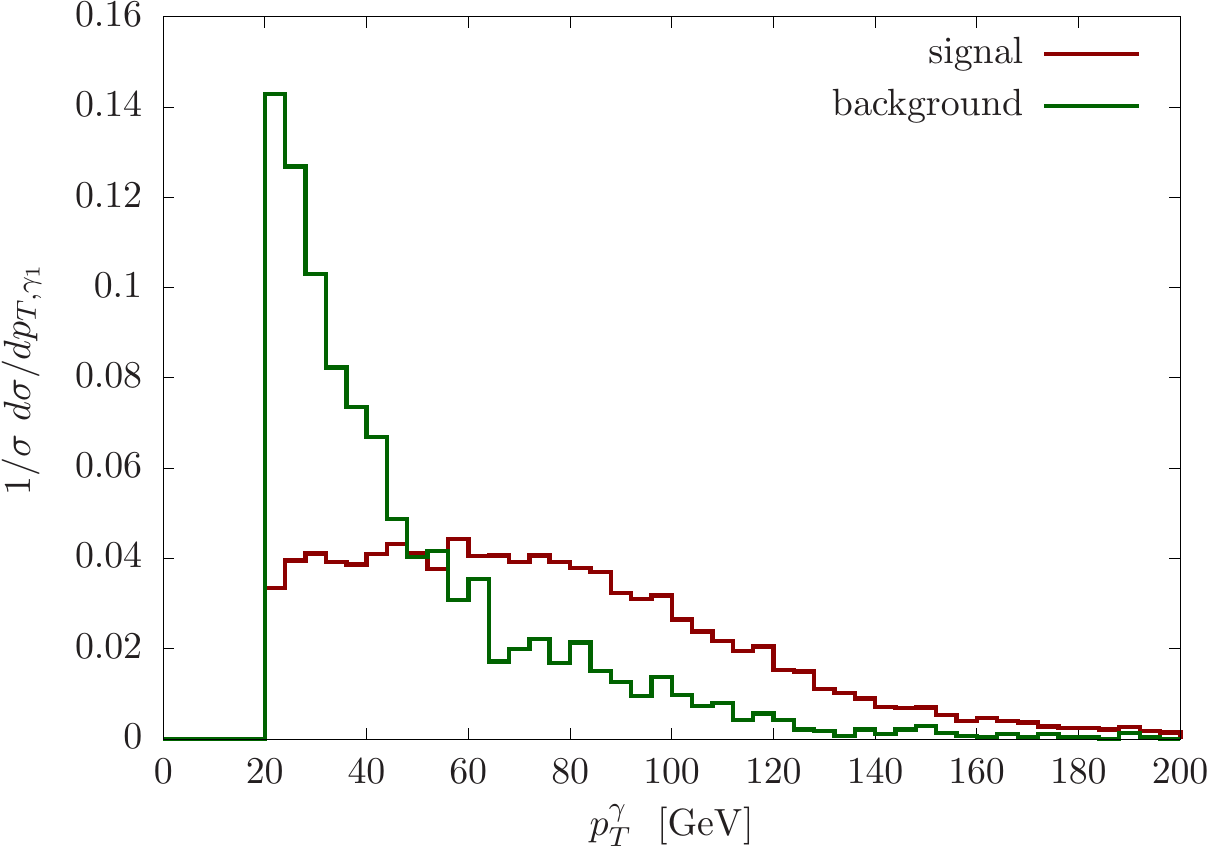}
\caption{$m_{\ell\ell\gamma} \equiv m_{H}$ (top) and $p^\gamma_T$ (bottom) distributions after event selection (for $m_{\ell\ell\gamma}$ the 
requirement $115$ GeV $\leq m_{\ell\ell\gamma} \leq 135$ GeV is not applied). Distributions are shown for three reference cuts of the transverse 
momentum of the leading jet: $p_T^{j_1} > 50$ GeV (left), $p_T^{j_1} > 100$ GeV (middle) and $p_T^{j_1} > 180$ GeV (right).}
\label{fig:hboson_ptj}
\end{figure}
  
\vspace{-1mm}  
  
\begin{table}[h!]
\begin{tabular}{| c | c | c | c | c | c |}
\hhline{~~----}
 \multicolumn{2}{ c | }{}   & $\sigma_S$ (fb)  & $\sigma_B$ (fb) & $S/B$ & $S/\sqrt{B}$ ($\mathcal{L} = 3000\,\,\mathrm{fb}^{-1}$) \\ [0.5ex]
\hline
 &Event Selection & 0.424 & 61.515 & 0.0069 & 2.96\\ [0.5ex]
\hhline{~-----}
$p^{j_{cut}}_T = 50$ GeV & $m_{\ell\ell\gamma} \in [122,128]$ GeV & 0.408 & 12.243 & 0.033 & 6.39\\ [0.5ex]
\hhline{~-----}
& $p^{\gamma}_{T} \geq 35$ GeV & 0.250 & 4.413 & 0.057 & 6.52 \\ [0.5ex]
\hline
&Event Selection & 0.131 & 10.14  & 0.0129 & 2.25 \\ [0.5ex]
\hhline{~-----}
$p^{j_{cut}}_T = 100$ GeV & $m_{\ell\ell\gamma} \in [122,128]$ GeV & 0.125  & 1.974 & 0.063 & 4.88 \\ [0.5ex]
\hhline{~-----}
&$p^{\gamma}_{T} \geq 40$ GeV  & 0.087  & 0.775 & 0.112 & 5.41 \\ [0.5ex]
\hline
& Event Selection & 0.034  & 1.446  & 0.0232  & 1.53 \\ [0.5ex]
\hhline{~-----}
$p^{j_{cut}}_T = 180$ GeV & $m_{\ell\ell\gamma} \in [122,128]$ GeV & 0.032  & 0.297  & 0.108 & 3.23 \\ [0.5ex]
\hhline{~-----}
&$p^{\gamma}_{T} \geq 45$ GeV & 0.024 & 0.120 & 0.203 & 3.86 \\ [0.5ex]
\hline
\end{tabular}
\caption{\small \small 13 TeV LHC cross section (in fb) for SM signal $\sigma_S$ and background $\sigma_B$ after event selection, 
Higgs mass window cut $m_{\ell\ell\gamma} \in [122,128]$ GeV and a further sliding cut on $p_T^{\gamma}$, respectively for 
a leading jet transverse momentum cut $p^{j_{cut}}_T = 50,\,100$, $180$ GeV. The values of signal-to-background $S/B$ and statistical 
significance $S/\sqrt{B}$ (for $\mathcal{L} = 3000\,\,\mathrm{fb}^{-1}$) at each stage of the analysis are also shown.}
\label{Table1}
\end{table}

\end{widetext}

Further SM background suppression may be achieved 
by imposing a tighter Higgs invariant mass window $m_{\ell\ell\gamma} \in [122,128]$ GeV
in accordance with~\cite{ATL-PHYS-PUB-2014-006}. 
As shown in Figure~\ref{fig:hboson_ptj}, this is more efficient for higher values of $p^{j_1}_T$. 
In addition, a harder cut on $p^{\gamma}_{T}$ correlated with the value of $p^{j_{cut}}_T$ yields a further
background reduction, as can be seen from Figure~\ref{fig:hboson_ptj} (bottom).

The results in Table~\ref{Table1} show that requiring the Higgs to be produced in association with a moderately boosted
jet, such that $p^{j_{cut}}_T \in [50,\,100]$ GeV, significantly improves both $S/B$ and $S/\sqrt{B}$ with respect to the 
inclusive search. The results also show that the potentially achievable value of $S/B$ increases with 
larger $p^{j_{cut}}_T$, however at the expense of a lower significance $S/\sqrt{B}$ due to the very small signal cross section at the LHC.
A high value $p^{j_{cut}}_T > 100$ GeV would then be particularly helpful if the measurement of $h \to Z \gamma$ at the HL-LHC 
is systematically limited.

 \begin{center}
\textit{(ii) Higgs production in weak boson fusion}
\end{center}
 
We now explore alternative Higgs production modes at the LHC which yield a fairly boosted Higgs boson, together with a not-so-small cross section.
These two requirements single out the WBF topology\footnote{We note that $h \to Z \gamma$ in WBF has already been 
initially studied in~\cite{ATL-PHYS-PUB-2013-014}.}, where 
the Higgs boson recoils against two highly energetic forward jets, 
as the only alternative to the $h +$jet Higgs production mode analyzed in the previous section.
Our signal corresponds to Higgs production in association with two jets, both in WBF and in gluon fusion (GF), while the relevant SM background
is $Z\gamma jj$ ($Z \to \ell^{+} \ell^{-}$).

\begin{widetext}
 \onecolumngrid
 
\begin{figure}[h!]
\centering
\includegraphics[width=0.325\textwidth]{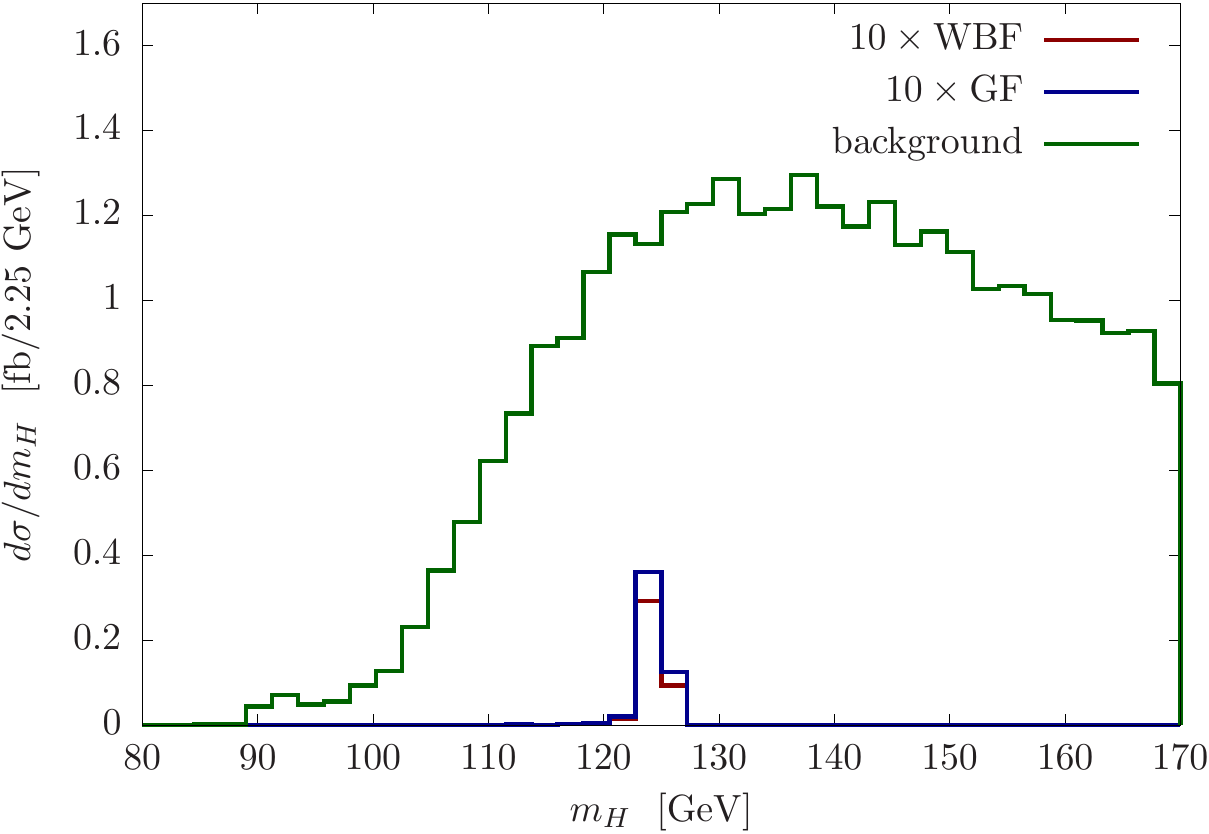}
\includegraphics[width=0.325\textwidth]{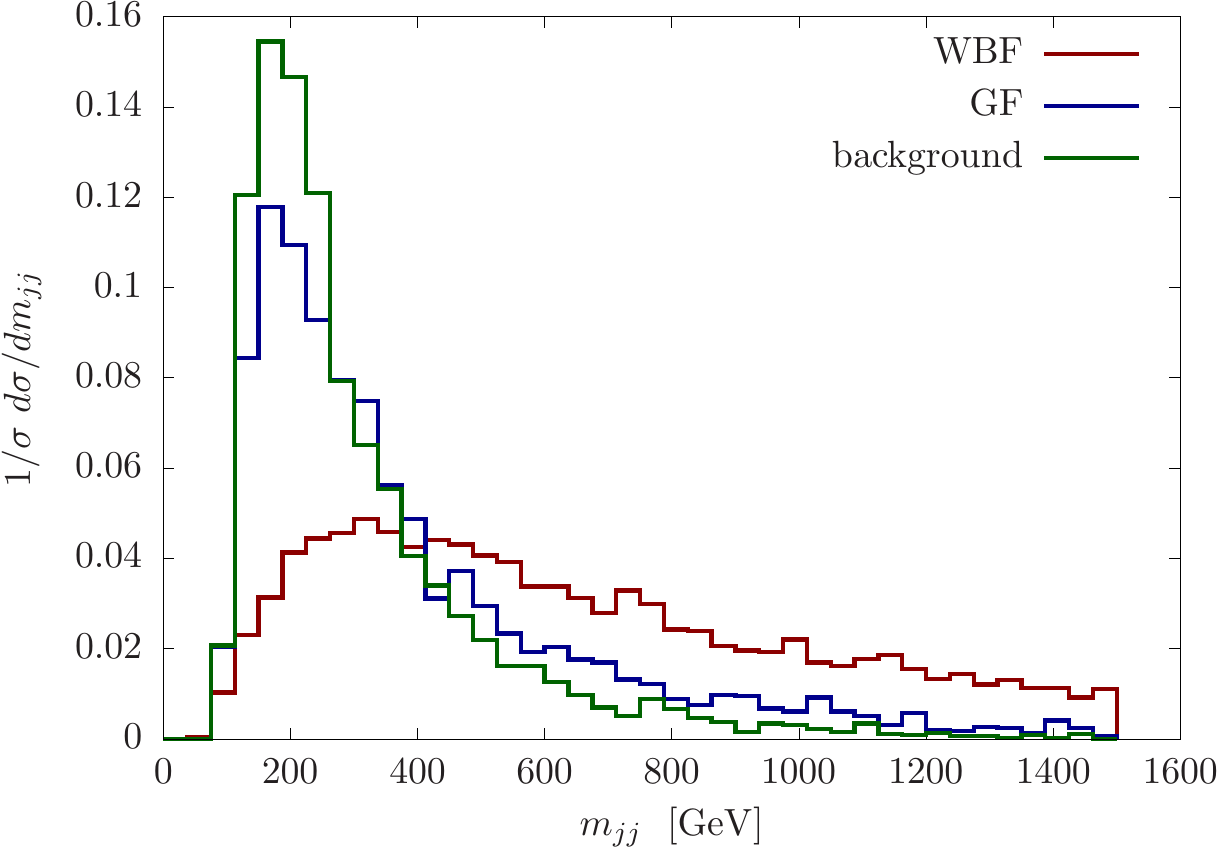} 
\includegraphics[width=0.325\textwidth]{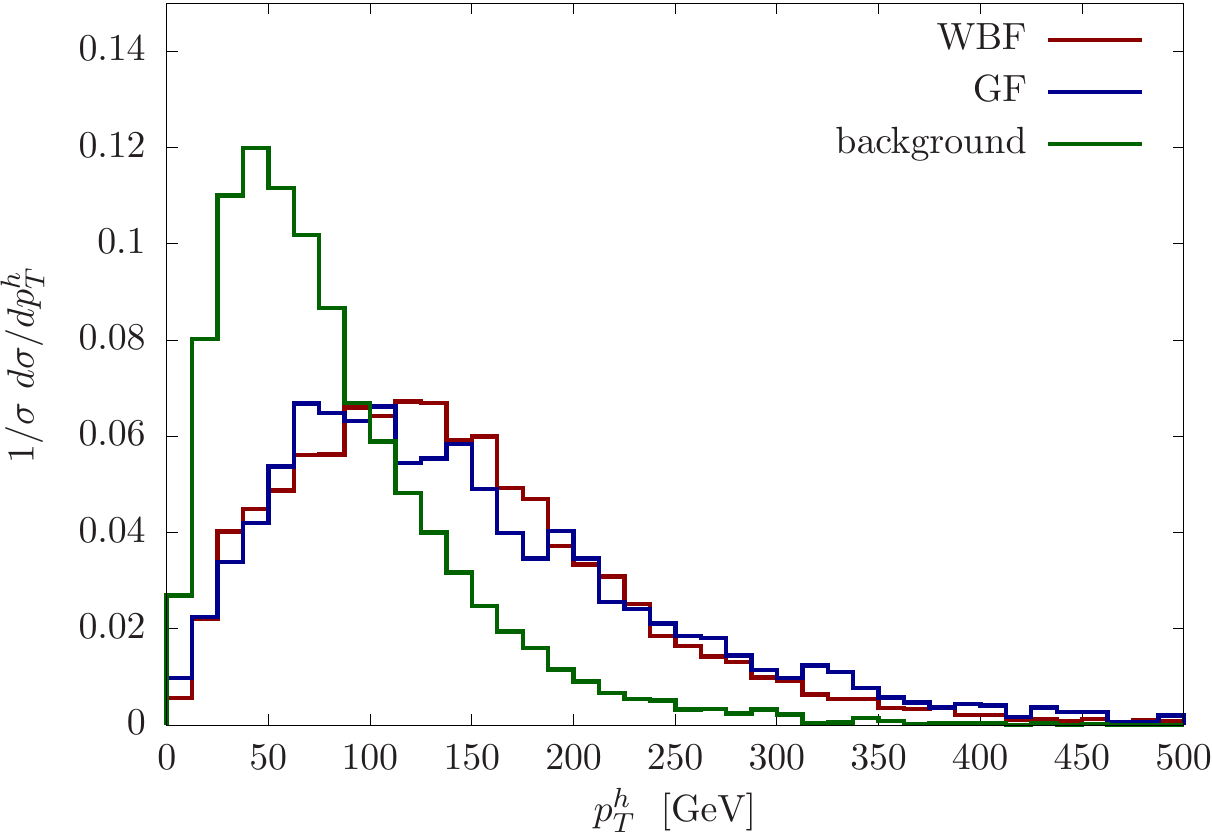} 
\caption{$m_{\ell\ell\gamma} \equiv m_{H}$ (left), $m_{jj}$ (middle) and  $p_T^{\ell\ell\gamma} \equiv p_{T}^h$ (right) 
distributions for signal (WBF and GF) and background after event selection (for $m_{\ell\ell\gamma}$ the 
selection $115$ GeV $\leq m_{\ell\ell\gamma} \leq 135$ GeV is not imposed).}
\label{fig:hboson_wbf}
\end{figure} 
 
 \vspace{-4mm}

\begin{table*}[h!]
\begin{tabular}{| c | c | c | c | c |}
\hline
 & Event Selection & $m_{\ell\ell\gamma} \in [122,128]$ GeV & $|\Delta y_{jj}| > 3.0$ 
 & $p^{h}_{T} > 80$ GeV \\
\hline
$\sigma_{\mathrm{WBF}}$ (fb) &  0.041  & 0.039 &  0.027 & 0.0208 \\ [0.5ex]
$\sigma_{\mathrm{GF}}$ (fb) &   0.051 & 0.049 & 0.011 & 0.0084 \\ [0.5ex]
$\sigma_{B}$ (fb) & 12.168  & 3.753  & 0.373 & 0.1512  \\ [0.5ex]
$S/B$ & 0.0076 & 0.0236 &  0.102 &  0.193 \\ [0.5ex]
$S/\sqrt{B}$ ($3000\,\,\mathrm{fb}^{-1}$) & 1.44  & 2.51 & 3.40 &  4.11  \\ [0.5ex]
\hline
\end{tabular}
\caption{\small 
13 TeV LHC cross section (in fb) in the $\ell^+ \ell^- \gamma j j$ final state for WBF signal $\sigma_{\mathrm{WBF}}$, GF signal $\sigma_{\mathrm{GF}}$ 
and background $\sigma_{B}$ after event selection, Higgs mass window cut $m_{\ell\ell\gamma} \in [122,128]$ GeV and further WBF selection cuts
$|\Delta y_{jj}| > 3.0$ and $p^h_{T} > 80$ GeV. The values of signal-to-background $S/B$ and statistical 
significance $S/\sqrt{B}$ (for $\mathcal{L} = 3000\,\,\mathrm{fb}^{-1}$) at each stage of the analysis are also shown.}
\label{Table3}
\end{table*} 
 
 
\end{widetext}

We adopt the same event generation and event selection criteria as in the previous 
Section, i.e. $p^{\ell}_T > 10$ GeV, $\left| y_{\ell} \right| < 2.5$, $p^{\gamma}_T > 20$ GeV, $\left| y_{\gamma} \right| < 2.5$, as well as the 
invariant mass windows
$
80~\mathrm{GeV} \leq m_{\ell\ell} \leq 100~\mathrm{GeV},
$
$
115~\mathrm{GeV} \leq m_{\ell\ell\gamma} \leq 135~\mathrm{GeV}.
$
For the  
$R=0.4$ anti-$kT$ jets we require $p^{j}_T \geq 50$ GeV, $|y_{j}| \leq 4.5$ and $\Delta R_{jj} \geq 2.0$.
Since NLO QCD corrections to weak boson fusion are known to be relatively small~\cite{Figy:2003nv, Cacciari:2015jma}, 
we adopt a $k$-factor $k=1.0$ for WBF, which we conservatively extend to the (subdominant after cuts) GF signal contribution.
Similarly, we use a flat NLO $k$-factor $k=1.2$ for the SM background~\cite{Campanario:2014wga}.

The $m_{jj}$, $p_T^{\ell\ell\gamma} \equiv p_{T}^h$ and $m_{\ell\ell\gamma}$ distributions
(the latter without imposing $115~\mathrm{GeV} \leq m_{\ell\ell\gamma} \leq 135~\mathrm{GeV}$) after event selection for WBF, GF and the background 
are shown in Figure~\ref{fig:hboson_wbf}.
To improve our signal sensitivity we further require a tight Higgs mass window $m_{\ell\ell\gamma} \in [122,128]$ GeV
as well as WBF selection cuts\footnote{After these two cuts, a potential $m_{jj} > 300$ GeV cut has only a mild impact on $S/B$.} 
$|\Delta y_{jj}| > 3.0$ and $p^{h}_{T} > 80$ GeV. 
The results for $S/B$ and $S/\sqrt{B}$ are shown in Table~\ref{Table3}, where it becomes clear that 
WBF does not perform as well as $h+$jet in terms of sensitivity to $h \to Z \gamma$. 
Still, a hypothetical combination of sensitivities between $h+$jet and WBF Higgs production (we perform a naive combination in quadrature)
could yield $S/\sqrt{B} \sim 7.7$, improving over the $h+$jet alone.
 
\vspace{-2mm}

\subsection{$h \to Z \gamma$ at Future $e^+e^-$ Colliders}

\vspace{-2mm}

We now explore the sensitivity to the decay $h \to Z \gamma$ that could be obtained for a future $e^+e^-$ collider. We focus on a circular
$e^{+}e^{-}$ collider (hereinafter {\sl fcc-ee}) in two configurations: a c.o.m.~energy of $\sqrt{s} = 240$ GeV, with a projected 
integrated luminosity 
of $10\,\, \mathrm{ab}^{-1}$~\cite{Gomez-Ceballos:2013zzn}, and a c.o.m.~energy of $\sqrt{s} = 350$ GeV, 
with a projected integrated luminosity of $2.6\,\, \mathrm{ab}^{-1}$~\cite{Gomez-Ceballos:2013zzn}. 
The Higgs production process considered is $e^{+} e^{-} \to Z h$, which has the highest cross section for an $e^+e^-$ collider with the chosen c.o.m. 
As compared to LHC, searches for $h \to (Z \to \ell^{+} \ell^{-})\gamma$ at {\sl fcc-ee} take advantage of the large amount of integrated 
luminosity as well as the presence of the extra $Z$ boson in the process. In order to maximize the sensitivity, we jointly consider the 
final states\footnote{The $Z \to \nu\bar{\nu}$ final state suffers from the large SM background $e^{+} e^{-} \to W^+ W^- \gamma$ ($W^+ W^- \to 2\ell\, 2 \nu$),
and thus we do not consider it here. Still, its addition could mildly improve the sensitivity to $h \to Z \gamma$.} 
$Z \to j j,\, \ell^+ \ell^-$ for the extra $Z$ boson. 
The dominant SM background is the irreducible $e^{+} e^{-} \to Z Z \gamma$, with $Z Z \to 2\ell\, 2j,\, 4\ell$.

\begin{widetext}
  \onecolumngrid
  
 \begin{figure}[h!]
\centering
\includegraphics[width=0.315\textwidth]{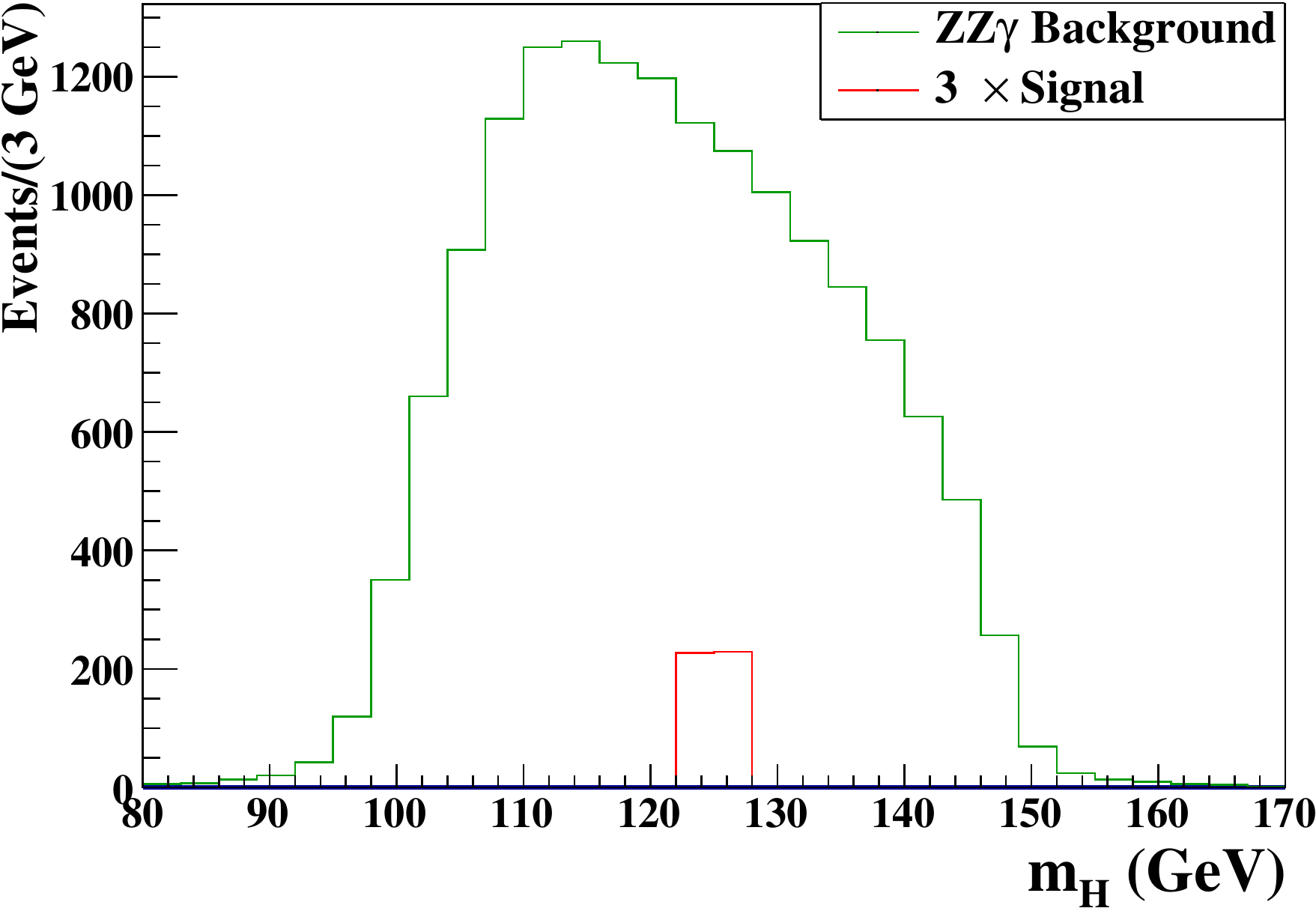}
\includegraphics[width=0.325\textwidth]{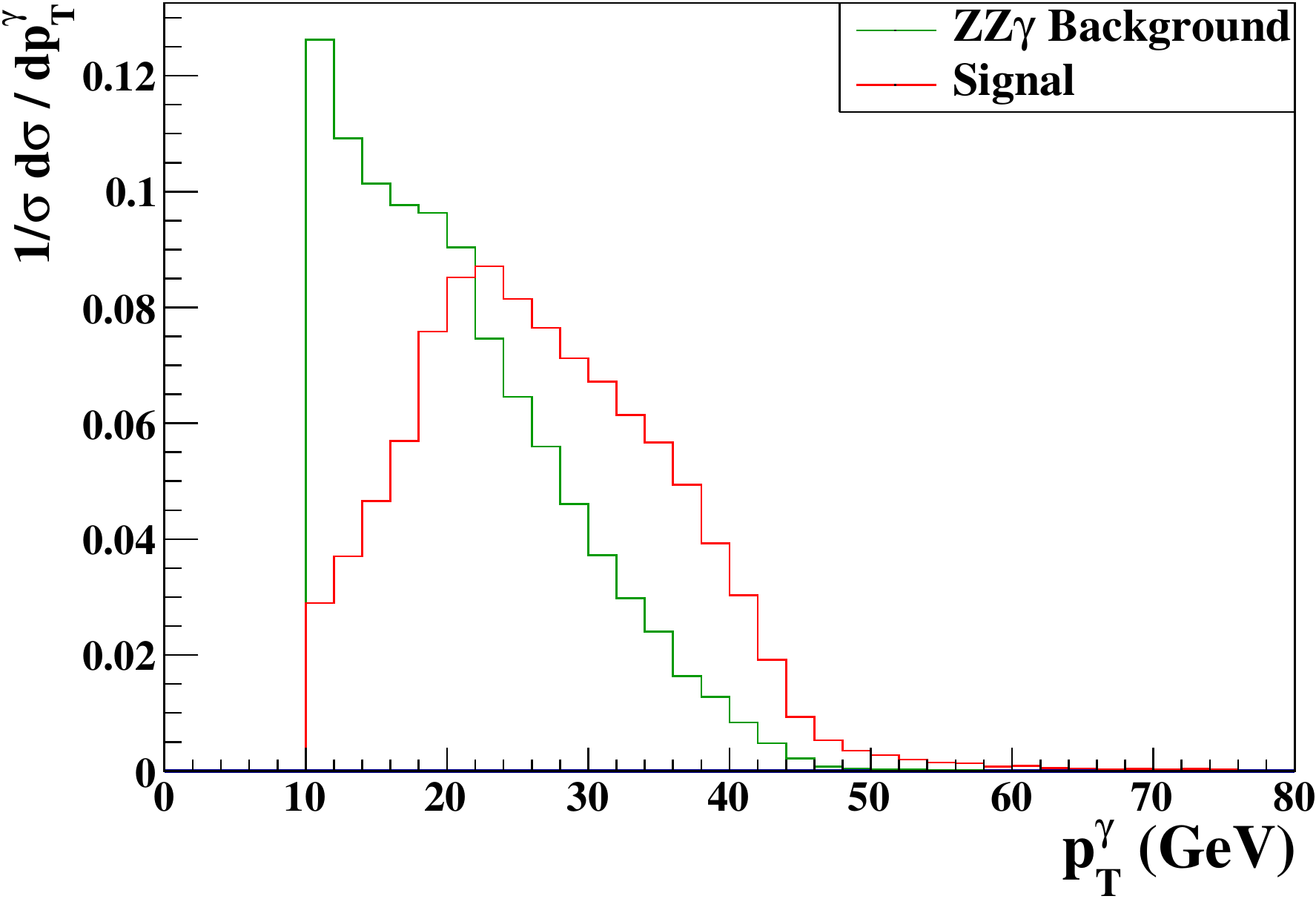} 
\includegraphics[width=0.325\textwidth]{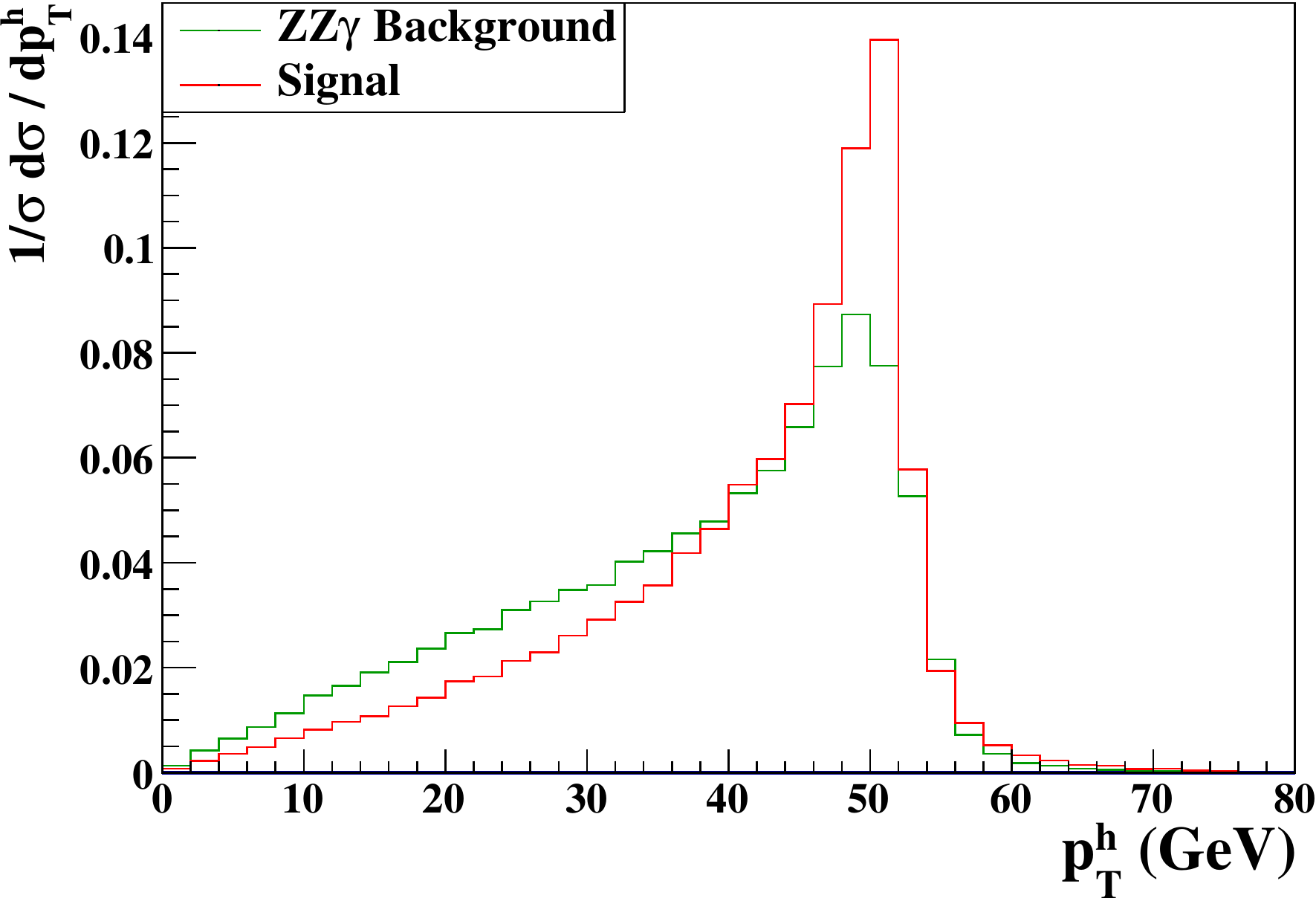} \\

\vspace{1mm}

\includegraphics[width=0.315\textwidth]{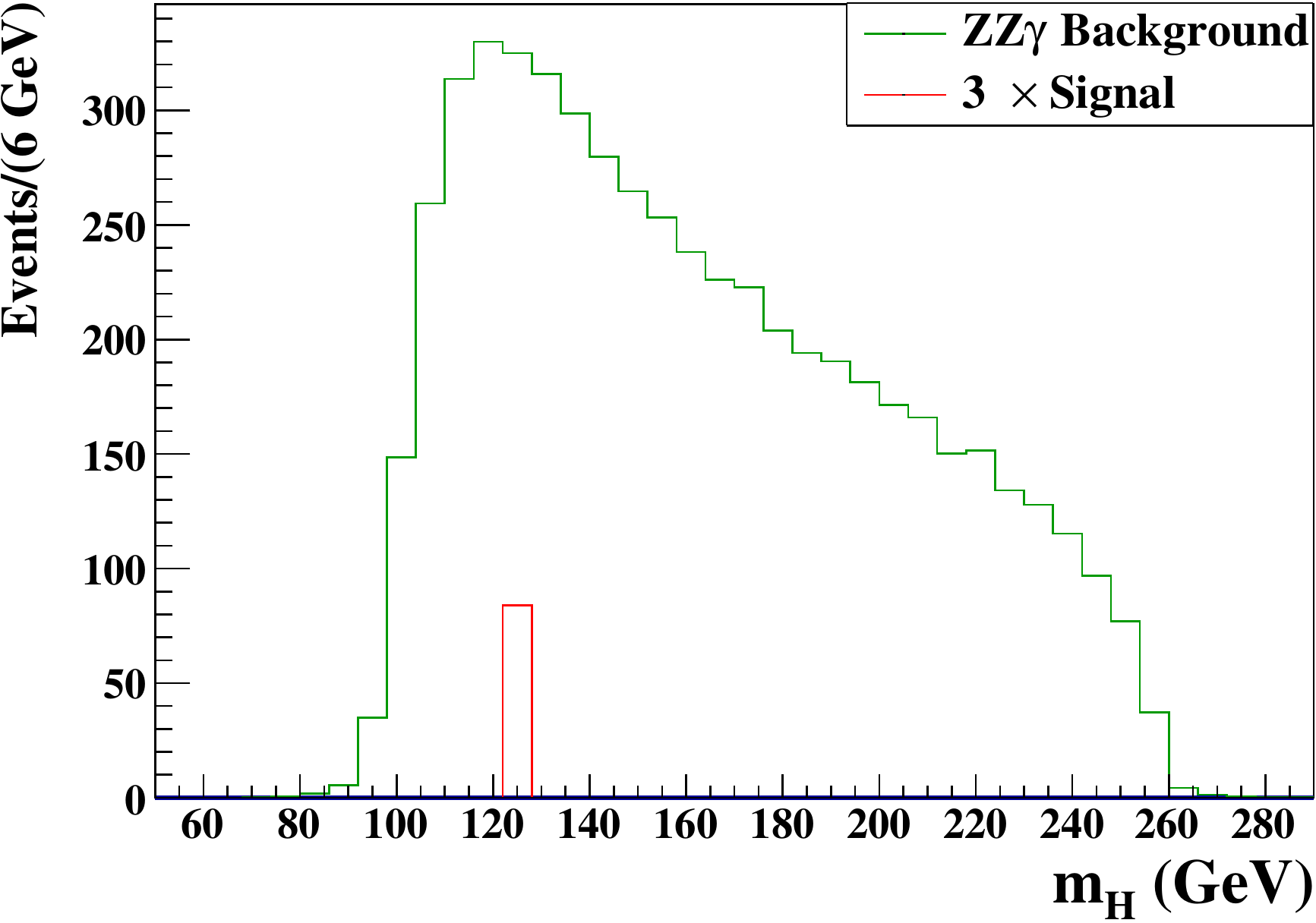}
\includegraphics[width=0.325\textwidth]{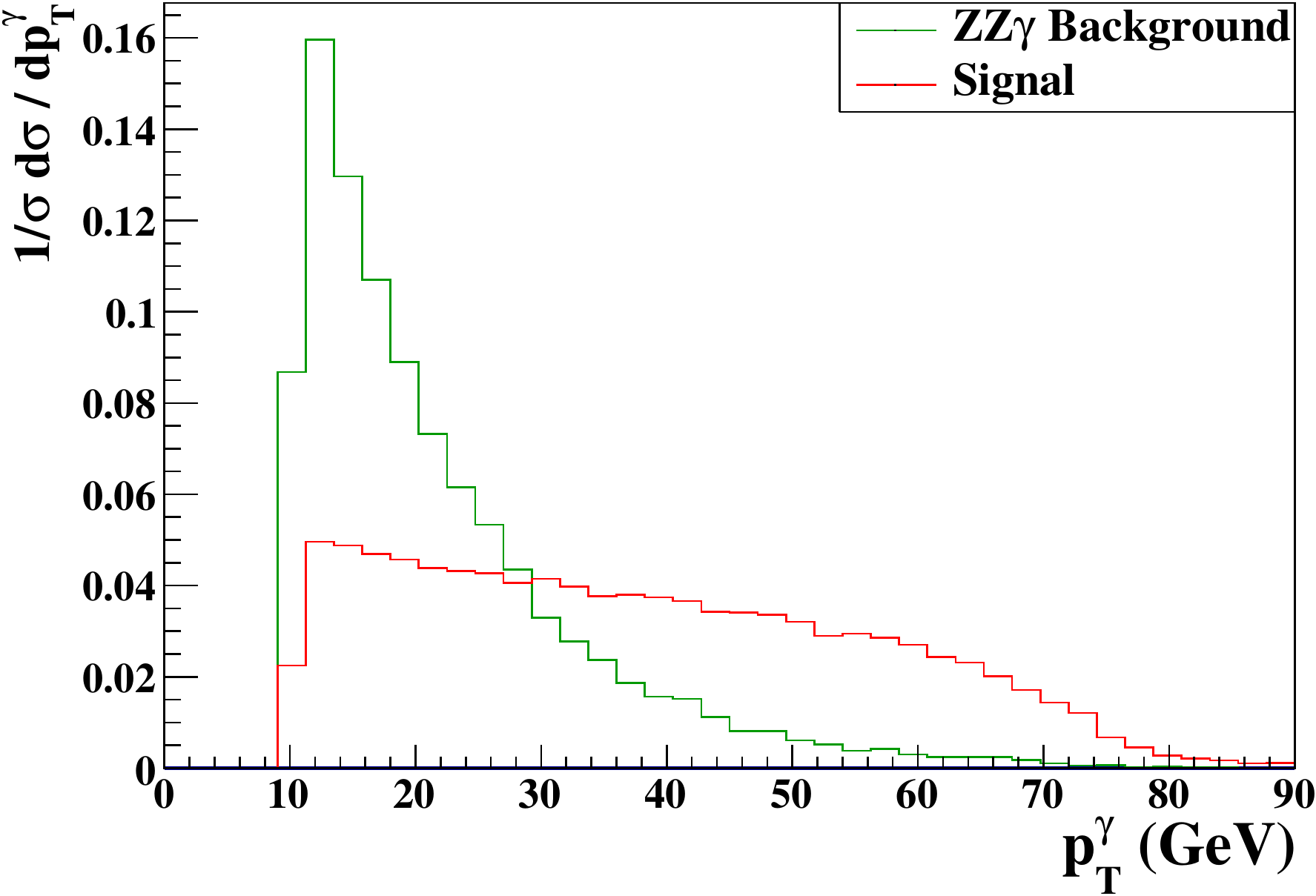}
\includegraphics[width=0.325\textwidth]{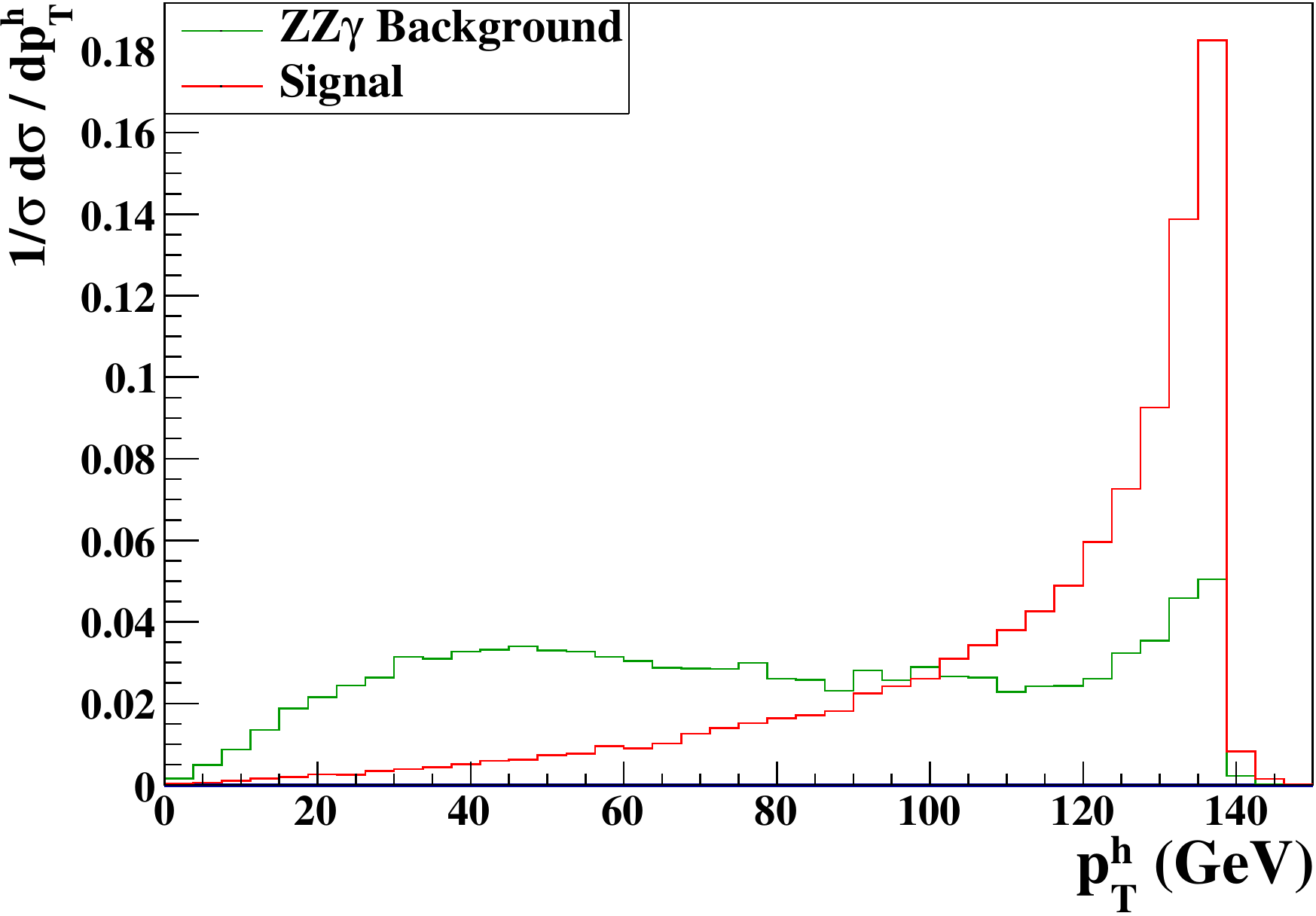} \\
\caption{Signal and background distributions for {\sl fcc-ee} $\sqrt{s} = 240$ GeV (top) and $\sqrt{s} = 350$ GeV (bottom) after event selection:
$m_{\ell\ell\gamma} \equiv m_{H}$ (left) without imposing the selection $115$ GeV $\leq m_{\ell\ell\gamma} \leq 135$ GeV.
$p^\gamma_T$ (middle) after requiring $m_{\ell\ell\gamma} \in [122,128]$ GeV. 
$p_T^{\ell\ell\gamma} \equiv p_{T}^h$ (right) after requiring $m_{\ell\ell\gamma} \in [122,128]$ GeV.
}
\label{fig:hboson_fccee}
\end{figure}

\vspace{-4mm}

\begin{table}[h!]
\begin{tabular}{| c | c | c | c | c | c |}
\hhline{~~----}
 \multicolumn{2}{ c | }{}   & $\sigma_S$ (fb)  & $\sigma^{ZZ\gamma}_B$ (fb) & $S/B$ & $S/\sqrt{B}$ \\ [0.5ex]
\hline
 &Event Selection & 0.0154 & 1.542 & 0.01 & 1.24\\ [0.5ex]
\hhline{~-----}
$\sqrt{s} = 240$ GeV & $m_{\ell\ell\gamma} \in [122,128]$ GeV & 0.0154 & 0.220 & 0.07 & 3.28\\ [0.5ex]
\hhline{~-----}
($\mathcal{L} = 10\,\, \mathrm{ab}^{-1}$) &$p^{\gamma}_{T} > 20$ GeV & 0.0116 & 0.103  & 0.113 & 3.62 \\ [0.5ex]
\hline
 &Event Selection & 0.0107 & 2.123  & 0.005 & 0.37 \\ [0.5ex]
\hhline{~-----}
$\sqrt{s} = 350$ GeV & $m_{\ell\ell\gamma} \in [122,128]$ GeV & 0.0107  & 0.125 & 0.086 & 1.54 \\ [0.5ex]
\hhline{~-----}
($\mathcal{L} = 2.6\,\, \mathrm{ab}^{-1}$) &$p^{\gamma}_{T} > 20$ GeV, $p^{h}_T > 90$ GeV  & 0.0071  & 0.020 & 0.354 & 2.56 \\ [0.5ex]
\hline
\end{tabular}
\caption{\small  {\sl fcc-ee} cross section for signal and background after event selection, 
Higgs mass window cut $m_{\ell\ell\gamma} \in [122,128]$ GeV and further selection cuts
$p_T^{\gamma} > 20$ GeV, $p^h_{T} > 90$ GeV. Values of $S/B$ and $S/\sqrt{B}$ 
at each stage of the analysis are also shown.}
\label{Table2}
\end{table}
\end{widetext}

We consider unpolarized $e^{+} e^{-}$ beams~\cite{Gomez-Ceballos:2013zzn,Mane:2014nqa}, and as in the previous Section we 
generate the signal and background with {\sc MadGraph\_aMC@NLO} and shower the partonic process 
with {\sc Pythia} 8. For event selection we require for the leptons $p^{\ell}_T > 10$ GeV, $\left| y_{\ell} \right| < 2.5$, for 
the photon $p^{\gamma}_T > 10$ GeV, $\left| y_{\gamma} \right| < 2.5$, and for the jets (for $Z \to j j$)
$p_{T}^j > 20$ GeV and $|y_j| < 5$, and further require $m_{\ell\ell} \in [80,\,\,100]$ GeV, $m_{jj} \in [80,\,\,100]$ GeV. 

The cross section for the signal $e^{+} e^{-} \to Z h$ is $\sigma_{Zh} = 0.193$ pb ($\sigma_{Zh} = 0.132$ pb) for $\sqrt{s} = 240$ GeV ($\sqrt{s} = 350$ GeV).
After the decay $h \to \ell^{+} \ell^{-} \gamma$ we then expect 
$\sim 154$ signal events with $\mathcal{L} = 10\,\, \mathrm{ab}^{-1}$ for $\sqrt{s} = 240$ GeV, and 
$\sim 28$ signal events with $\mathcal{L} = 2.6\,\, \mathrm{ab}^{-1}$ for $\sqrt{s} = 350$ GeV, 
which means that tight cuts are not helpful in extracting the signal due to the very small cross section. 
Cross sections for the SM signal and background after event selection are shown in Table~\ref{Table2}. 
As in the previous Section, we define our Higgs mass signal region as $m_{\ell\ell\gamma} \in [122,128]$ GeV. In order to further
increase $S/B$ in the signal region we require 
$p^{\gamma}_T > 20$ GeV (for $\sqrt{s} = 240,\,350$ GeV), together 
with $p^{h}_T > 90$ GeV (only for $\sqrt{s} = 350$ GeV). 
The invariant mass distribution  $m_{\ell\ell\gamma}$ as well as the 
$p^{\gamma}_T$ and $p^{h}_T$ distributions in the signal region are shown in Figure~\ref{fig:hboson_fccee}.  

The results in Table~\ref{Table2} show that, while the values of $S/B$ achievable by a future 
{\sl fcc-ee} machine in the measurement of $h \to Z \gamma$ may be higher than those of LHC (for 
$\sqrt{s} = 350$ GeV), a precise measurement at {\sl fcc-ee} is limited by statistics, and the projected signal 
significance at the HL-LHC in the $h+$jet production channel is significantly larger.

\vspace{-2mm}

\subsection{Probing New Physics with $h \to Z \gamma$}

\vspace{-2mm}

Precisely measuring the $125$ GeV Higgs boson signal strengths in the $h\to Z\gamma$ and $h\to\gamma\gamma$ final states could reveal 
the existence of new charged particles coupled to $h$ (see e.g.~\cite{Chiang:2012qz,Carena:2012xa,Bizot:2015zaa}). While the results of the above sections 
show that a precise measurement of 
the signal strength for $h \to Z \gamma$ is possible at the LHC (and to a lesser extend, at a future {\sl fcc-ee} collider), 
the achieved precision in the measurement of the $h\to\gamma\gamma$ signal strength will be much higher in both colliders, and as such 
new charged particles coupled to the Higgs would first manifest themselves via a deviation in the $h\to\gamma\gamma$
channel w.r.t. the SM value.
Still, the precise measurement of the $h\to Z\gamma$ signal strength would yield valuable complementary information to the 
$h\to\gamma\gamma$ channel, regarding the $SU(2)_{\mathrm{L}} \times U(1)_{\mathrm{Y}}$ quantum numbers of these new charged particles.

Let us consider the partial widths $\Gamma (h \to \gamma \gamma) \equiv \Gamma_{\gamma\gamma}$ and $\Gamma (h \to Z \gamma) \equiv \Gamma_{Z\gamma}$ in the presence of 
a BSM contribution encoded in the effective operators
\begin{equation}
\label{eff_operators}
\kappa_B \frac{\alpha_{\mathrm{EM}}}{8\pi v\, c_{W}^2} B_{\mu\nu} B^{\mu\nu} \quad , \quad 
\kappa_W \frac{\alpha_{\mathrm{EM}}}{8\pi v\, s_{W}^2} W^a_{\mu\nu} W_a^{\mu\nu} \,.
\end{equation}
with $s_W$ ($c_W$) being the sine (cosine) of the Weinberg angle. The partial widths read

\begin{widetext}
 \begin{eqnarray}
\Gamma_{\gamma\gamma} = \frac{\alpha^2_{\mathrm{EM}} m_h^3}{256 \,\pi^3 v^2} \left| \kappa_W + \kappa_B - F_1(\tau_W) - \sum_f 3 Q^2_f F_{1/2}(\tau_f) \right|^2\\
\Gamma_{Z\gamma} = \frac{\alpha^2_{\mathrm{EM}} m_h^3}{128 \,\pi^3 v^2} \left(1- \frac{m_Z^2}{m_h^2} \right)^3 
\left| t^{-1}_W\kappa_W - t_W \kappa_B - A_W(\tau_W,\lambda_W) - \sum_f 3 \frac{Q_f (2 I_f^3 - 4 Q_f s_W^2)}{c_W} A_{f}(\tau_f,\lambda_f) \right|^2
\end{eqnarray}
\end{widetext}
with $Q_f$ and $I^3_f$ the electric charge and third component of weak isospin of the SM fermions $f$ entering the 
$h\to \gamma\gamma$ and $h\to Z\gamma$ loops. The form factors $F_1(x)$, $F_{1/2}(x)$, $A_W(x,y)$, $A_{f}(x,y)$ are 
given in~\cite{Gunion:1989we} (see also~\cite{Fontes:2014xva}), with $\tau_i = 4 m_i^2 /m_h^2 $ and $\lambda_i =  4 m_i^2 / m_Z^2$. 

For $h \to \gamma \gamma$, Higgs signal strength measurements with 7 - 8 TeV LHC data yield a signal strength value 
$\mu_{\gamma\gamma} = 1.17 \pm 0.27$~\cite{Aad:2015gba}. 
For 14 TeV LHC with 3000 fb$^{-1}$ the projected signal strength sensitivity is
$\Delta \mu_{\gamma\gamma} = 0.04$~\cite{ATL-PHYS-PUB-2014-016}.
From the present analysis, for the HL-LHC with 3000 fb$^{-1}$ the projected signal strength sensitivity
in $h \to Z \gamma$ (combining the results from $h+$jet and WBF Higgs production) 
is $\Delta \mu_{Z\gamma} \simeq 0.13$ neglecting 
systematic uncertainties ($\Delta \mu_{Z\gamma}  \simeq 0.2$ considering 
the same amount of systematics as in~\cite{ATL-PHYS-PUB-2014-006}).
In order to illustrate the potential gain of studying the decays $h \to \gamma\gamma$ and $h \to Z \gamma$ in concert, 
let us consider two alternative hypothetical scenarios:

\vspace{1mm}

\noindent {\textit{(i)} No deviation in $\mu_{\gamma\gamma}$ and $\mu_{Z\gamma}$ w.r.t.~to the SM is measured at the HL-LHC.

\vspace{1mm}

\noindent {\textit{(ii)} A deviation in $\mu_{\gamma\gamma}$ w.r.t.~to the SM is measured at the HL-LHC. 
We choose $\mu_{\gamma\gamma} = 1.17$ (corresponding to the LHC Run 1 central value for the signal strength),
and consider three possible measured values $\mu_{Z\gamma} = 0.5,\,0.65,\,0.8$ at the HL-LHC.

\vspace{1mm}

In these two scenarios, we then show in Figure~\ref{Fit_Zgamma} the would-be 95 \% C.L. limits in the ($\kappa_W$, $\kappa_B$) plane
from a $\chi^2$ fit (assuming for simplicity no systematic uncertainties in the HL-LHC measurement of
$\mu_{Z\gamma}$). 
For scenario {\it (i)} (Figure~\ref{Fit_Zgamma}, Top) the combination of $\mu_{\gamma\gamma}$ and $\mu_{Z\gamma}$ constrains 
a blind direction (present for $\mu_{\gamma\gamma}$ only) in the ($\kappa_W$, $\kappa_B$) plane, limiting 
$\kappa_B$ to the range $[-1.42, \,1.58]$ and $\kappa_W$ to the range $[-1.32,\,1.17]$ at 95 \% C.L. We also show for comparison the corresponding would-be
constraint (dashed-blue) from an inclusive measurement of $\mu_{Z\gamma}$ with projected incertainty $\Delta \mu_{Z\gamma} \sim 0.3$~\cite{ATL-PHYS-PUB-2014-016}.  
For scenario {\it (ii)} (Figure~\ref{Fit_Zgamma}, Bottom), the combination of $\mu_{\gamma\gamma}$ and $\mu_{Z\gamma}$
allows to extract bounds on the $SU(2)_{\mathrm{L}} \times U(1)_{\mathrm{Y}}$ properties of the would-be 
charged particle(s) responsible for the signal strength deviations. 
The ratio $\kappa_B/\kappa_W$ is given by
\begin{equation}
\label{kappa}
\frac{\kappa_B}{\kappa_W} \equiv \mathrm{tan}(\Theta) = \frac{12\,\mathrm{Y}^2}{(N-1)(N+1)} 
\end{equation}
where Y is the hypercharge of the new charged particle and $N = 1,\,2,\,3,\,4...$ denotes its $SU(2)_{\mathrm{L}}$ representation
($1 =$ singlet, $2 =$ doublet, $3 =$ triplet...). The notation $\kappa_B/\kappa_W = \mathrm{tan}(\Theta)$ has been introduced for 
convenience (see e.g.~\cite{Sato:2016hls}). The measurements $\mu_{Z\gamma} = 0.5,\,0.65,\,0.8$ respectively yield the bounds 
$\Theta \in [2.39,\,3.07]$, $\Theta \in [2.40,\,4.55]$, $\Theta \in [2.43,\,5.22]$ at 95 \% C.L. In the first two cases the $SU(2)_{\mathrm{L}}$ singlet 
hypothesis ($\Theta = \pi/2,\,3\pi/2$) would be disfavoured at more than $2\sigma$ by means of the combined $\mu_{Z\gamma}$, $\mu_{\gamma\gamma}$ measurement, 
while in the first case the $\mathrm{Y} = 0$ hypothesis ($\Theta = 0,\,\pi$) would also be disfavoured at more than $2\sigma$.

\begin{figure}[t!]
\centering
\includegraphics[width=0.45\textwidth]{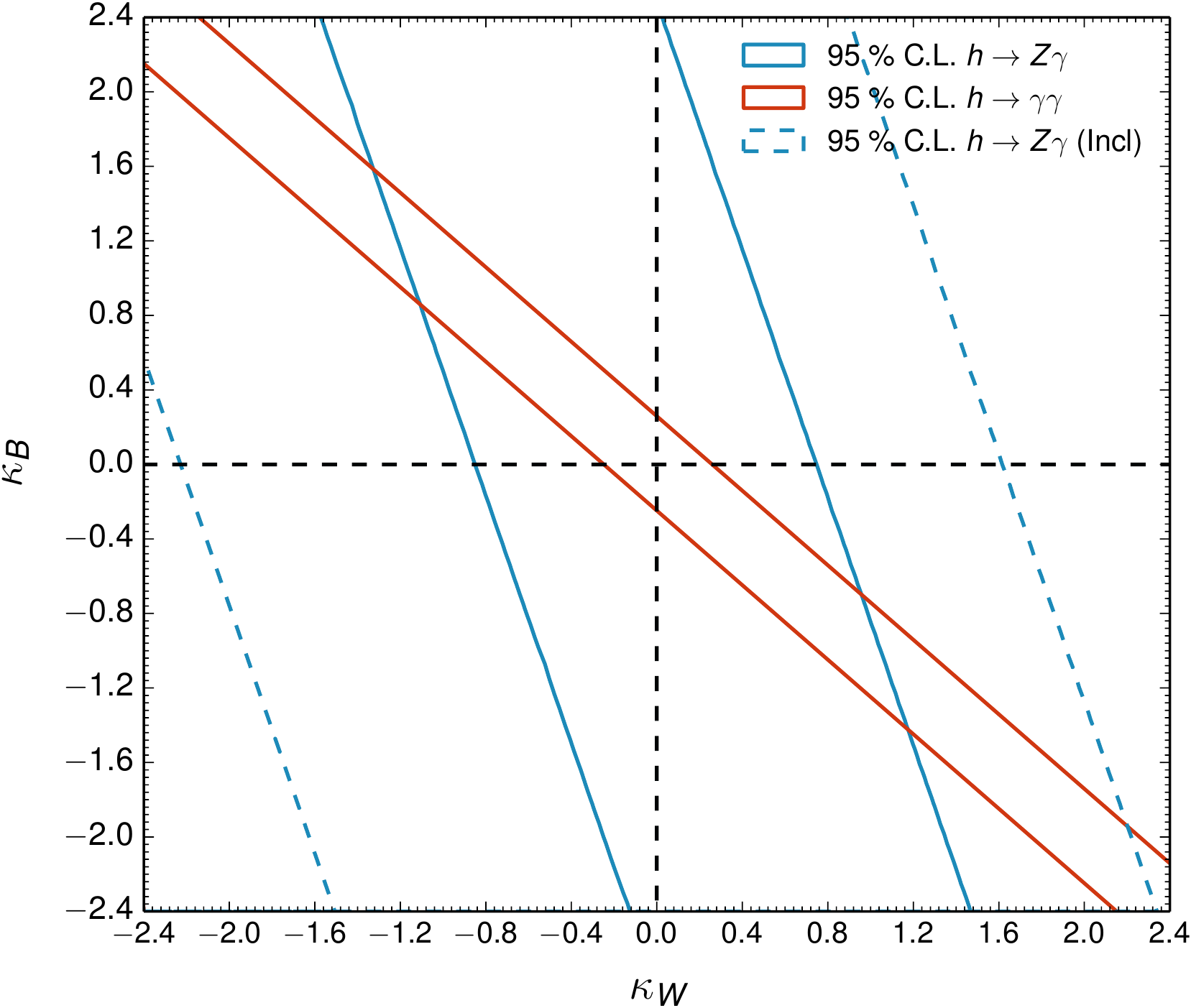}
\includegraphics[width=0.45\textwidth]{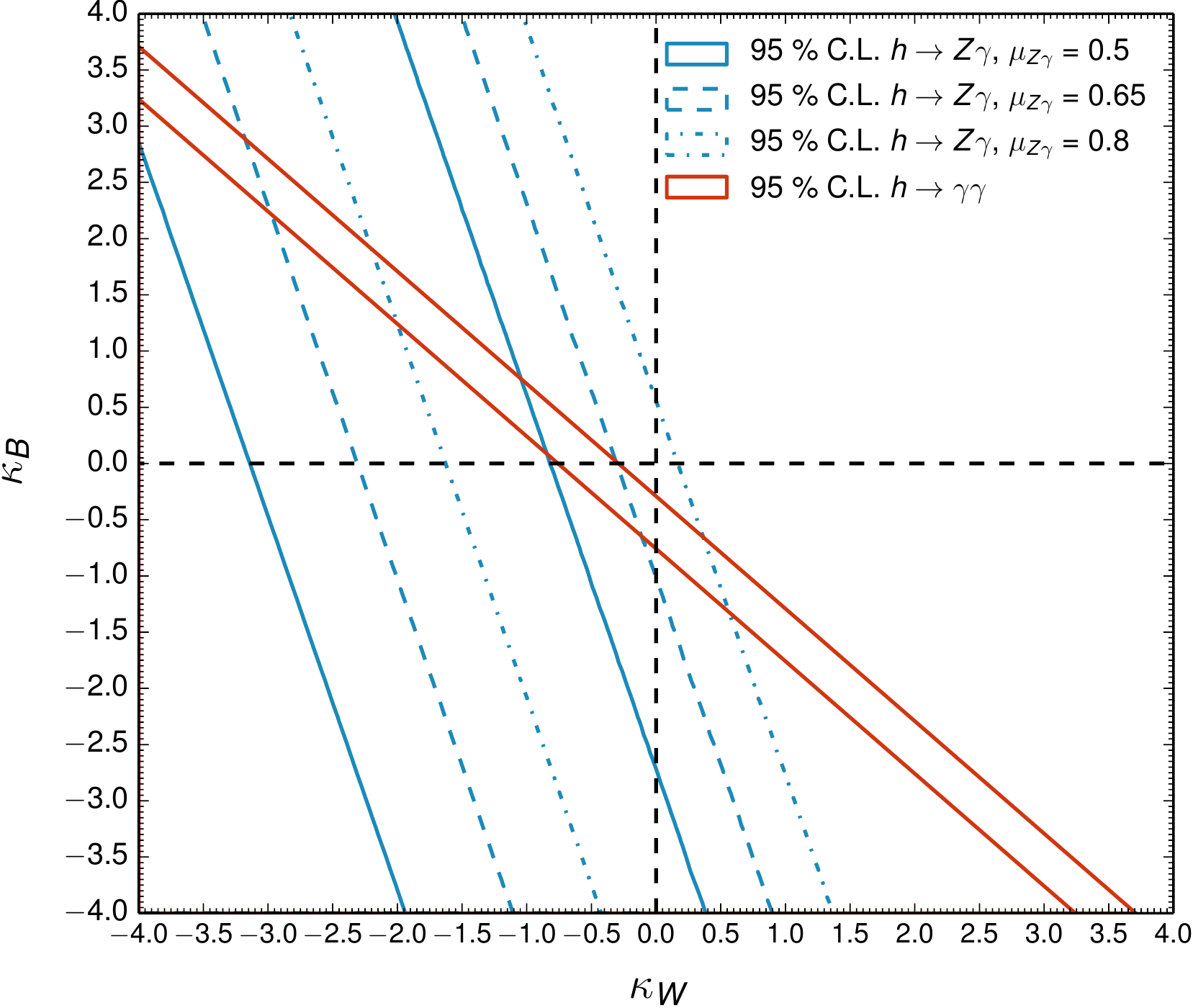}
\caption{\small 95 $\%$ C.L. bounds in the ($\kappa_W$, $\kappa_B$) plane from measurements of Higgs signal strengths $\mu_{\gamma\gamma}$ (red) and 
$\mu_{Z\gamma}$ (blue) at HL-LHC. Top: Assuming measured central values $\mu_{\gamma\gamma} = 1$, $\mu_{Z\gamma} = 1$ (no deviations from SM). 
For, $\mu_{\gamma\gamma}$, we show the comparison of the present analysis (solid) with an inclusive measurement (dashed).  Bottom:
Assuming measured central values $\mu_{\gamma\gamma} = 1.17$ and $\mu_{Z\gamma} = 0.5,\,0.65,\,0.8$, respectively solid, dashed, dot-dashed.}
\label{Fit_Zgamma}
\end{figure}

\vspace{3mm}

Before concluding, it is also important to stress that the $h \to Z \gamma$ decay is also 
a potential window into new physics in other contexts: 
Interference effects involving $h \to Z \gamma$ have been studied in~\cite{Korchin:2013ifa,Chen:2014gka,Chen:2014ona,Farina:2015dua} as a future probe of the 
CP properties of the 125 GeV Higgs boson, and improving the sensitivity to 
$h \to Z \gamma$ could also improve some of these measurements.
%
Similarly, precise measurements of $h \to Z \gamma$ would also allow to constrain 
the presence of light pseudoscalars $a$ to which the Higgs can decay (via $h \to Z \,a$), 
for a dominant pseudoscalar decay $a \to \gamma\gamma$ (see e.g.~\cite{Mimasu:2014nea,Jaeckel:2015jla}) with two very collimated 
photons (which do not get resolved in the detector).

\vspace{-2mm}

\subsection{Conclusions}

\vspace{-2mm}

The decay $h \to Z \gamma$ is very challenging to measure precisely at the LHC or a future 
$e^{+}e^{-}$ collider. In this work we have shown that considering Higgs production in association with a moderately
boosted jet at the LHC can significantly improve the sensitivity in this channel compared to the inclusive measurement,
due to kinematic (de)correlations among $p_T^{j}$, $m_{\ell\ell\gamma}$ and $p_T^{\gamma}$ 
for signal and background in the $\ell\ell\gamma+j$ final state. By combining these measurements with those of 
$h \to Z\gamma$ in the (less sensitive) weak boson fusion Higgs production channel, we show that it could be possible
to achieve $\Delta \mu_{Z\gamma} \sim 0.13$ (in the absence of systematic uncertainties) at the HL-LHC with
$\mathcal{L} = 3000$ fb$^{-1}$. We have also illustrated the potential of such a measurement to probe 
new physics scenarios, particularly to gain information on the $SU(2) \times U(1)$ quantum numbers of would-be charged 
particles coupled to the Higgs. 

Finally, we want to stress that the $h \to Z \gamma$ measurement strategy in $h+$jet discussed here, while 
yielding fair precision at the HL-LHC, could be optimally exploited at a 100 TeV proton-proton collider ({\sl fcc-pp}) (see {\it e.g.} the discussion in Section
4.2.3 of~\cite{Contino:2016spe}) due to the much higher $h+$jet cross section (particularly at high $p_T^{j}$), allowing for a
very strong background suppresion without being statistically limited by the small LHC cross sections.
We leave a detailed study of $100$ TeV prospects for $h \to Z \gamma$ and its phenomenological implications for the future.

\vspace{2mm}

\begin{center}
\textbf{Acknowledgements} 
\end{center}

\vspace{-1mm}

\begin{acknowledgments}
J.M.N. is supported in part by the People Programme (Marie Curie Actions) of the European Union Seventh
Framework Programme (FP7/2007-2013), REA grant agreement
PIEF-GA-2013-625809 and the European Research Council under the
European Union’s Horizon 2020 program (ERC Grant Agreement no.648680 DARKHORIZONS). 
M.S. is supported in part by the European Commission through 
the ``HiggsTools'' Initial Training Network PITN-GA-2012-316704.
\end{acknowledgments}


\end{document}